\documentclass[twocolumn,showpacs,preprintnumbers]{revtex4}
\usepackage{amsmath,amssymb}
\usepackage{graphicx}
\usepackage{dcolumn}
\usepackage[vcentermath]{youngtab}
\usepackage[dvips]{epsfig,color}
\usepackage{bm}
\begin{document}
\title{Higher mass bottomonia}
\author{J. Ferretti}
\affiliation{INFN, Sezione di Genova, via Dodecaneso 33, 16146 Genova (Italy)}
\affiliation{Instituto de Ciencias Nucleares, Universidad Nacional Aut\'onoma de M\'exico, 04510 M\'exico DF, M\'exico}
\author{E. Santopinto}\thanks{Corresponding author: santopinto@ge.infn.it}
\affiliation{INFN, Sezione di Genova, via Dodecaneso 33, 16146 Genova (Italy)} 
\begin{abstract}
We show the results of a coupled-channel quark model calculation of the bottomonium spectrum with self energy corrections due to the coupling to the meson-meson continuum.
We also provide results for the open bottom strong decay amplitudes of higher bottomonia in a $^3P_0$ model.
\end{abstract}
\maketitle

\section{Introduction}
\label{Introduction} 
Several studies on meson spectroscopy have been published over the years, based on different pictures for mesons; these include $q \bar q$ 
mesons \cite{Eichten:1974af,Quigg:1979vr,Buchmuller:1980su,Moxhay:1983vu,Godfrey:1985xj,Gupta:1986xt,Fulcher:1990kx,
Lucha:1991vn,Eichten:1994gt,Kiselev:1994rc,Godfrey:2004ya,Vijande:2007ke,Danilkin:2010cc,Radford:2009qi,Colangelo:2012xi,Dib:2012vw,
Segovia:2012cs}, 
comprising the unquenched lattice QCD calculations of Refs. \cite{Bali:1997am,Davies:1998im,Gray:2005ur,Burch:2007fj,Meinel:2010pv}, 
meson-meson molecules \cite{Hanhart:2007yq,Baru:2011rs,Aceti:2012cb,Weinstein:1990gu,Barnes:1991em,Tornqvist:1993ng,Swanson:2003tb}, 
tetraquarks \cite{Jaffe:1976ig,Maiani:2004uc,Santopinto:2006my} and quarkonium hybrids 
\cite{LlanesEstrada:2000hj,Buisseret:2006wc,Guo:2008yz}. Further references can be found in review papers \cite{REVIEW}. 

Many of these studies have investigated hadron properties within the quark model (QM).
The QM \cite{Eichten:1974af,Buchmuller:1980su,Moxhay:1983vu,Godfrey:1985xj,Gupta:1986xt,Fulcher:1990kx,Lucha:1991vn,
Colangelo:2012xi,Isgur:1979be,Capstick:1986bm,Glozman-Riska,Loring:2001kx,Giannini:2001kb,
Ferretti:2011,Valcarce:2005rr,Galata:2012xt} can reproduce the behavior of observables such as the spectrum and the magnetic moments, but 
it neglects pair-creation effects, which are manifested as a coupling to meson-meson (meson-baryon) channels.
Above threshold, this coupling leads to strong decays; below threshold, it leads to virtual $q \bar q - q \bar q$ ($qqq - q \bar q$) 
components in the hadron wave function and shifts of the physical mass in relation to the bare mass, as already shown by several 
authors in the meson \cite{Danilkin:2009hr,Kalashnikova:2005ui,Ono:1983rd,Barnes:2007xu,Hwang:2004cd,
Eichten:2004uh,Rupp:2006sb,Pennington:2007xr,Li:2009nr,Liu:2011yp,charmonium} and baryon \cite{Tornqvist,Horacsek:1986fz,Blask:1987yv,
Brack:1987dg,SilvestreBrac:1991pw,Fujiwara:1992yv,Morel:2002vk} sectors. 
Indeed, since the earliest days of hadron spectroscopy, it has been recognized that the properties of a level can be strongly influenced 
by the closest channels \cite{Patrignani:2012an,Burns:2012pc}. 
An early example was the resonance $\Lambda(1405)$, decaying into $\Sigma \pi$ but strongly influenced by the nearby $\bar K N$ threshold 
\cite{Dalitz:1959dn}, or the $f_0(980)$, decaying into $\pi \pi$ but behaving remarkably like a $K \bar K$ meson-meson molecule 
\cite{Flatte:1972rz}. 
The unquenching of the quark model for hadrons is a way to take into account pair-creation (or continuum) effects (i.e. meson-meson and 
baryon-meson configurations in the meson and baryon sectors, respectively).  

Pioneering work on the unquenching of meson quark models (one of the major developments in the field of QM's) was done by 
T\"ornqvist and collaborators \cite{Ono:1983rd,Tornqvist}, while van Beveren and Rupp used an heuristic t-matrix approach
\cite{vanBeveren:1979bd,vanBeveren:1986ea}.
This difficult method, sometimes also called unitarized quark model \cite{Ono:1983rd,Tornqvist}, was developed by T\"ornqvist and then also 
applied to the study of the scalar meson nonet ($a_0$, $f_0$, etc.) \cite{Tornqvist:1995kr}. The method was used (with a few variations) by 
T\"ornqvist, Isgur, Pennington, Bijker and Santopinto to study the influence of the meson-meson (meson-baryon) continuum on  meson (baryon) 
observables, such as the strangeness content of the nucleon e.m. form factors \cite{Bijker:2012zza}, the flavor asymmetry of the proton 
\cite{Santopinto:2010zza}, the importance of the orbital angular momentum in the spin of the proton, as explicitly calculated for the first 
time with a quark model in Ref. \cite{Bijker:2009up}, and the charmonium spectrum and threshold effects on the mass of the $X(3872)$ 
resonance \cite{charmonium,Ono:1983rd,Pennington:2007xr,Liu:2011yp}. 
The more heuristic and hybrid t-matrix approach developed by van Beveren and Rupp \cite{vanBeveren:1979bd,vanBeveren:1986ea}, with a quark 
model seed and a meson cloud, does not have wave functions, unlike the UQM. 
In any case, the two approaches, by means of wave functions or t-matrix, should be linked in the end, though a complete and difficult 
derivation from the first to the second is still to be worked out in detail, and will be the subject of a subsequent paper, in which the 
present approach will be extended to cross section calculations. 

The main aim of the present study is to calculate the bottomonium spectrum with self energy corrections, which requires the 
use of the UQM formalism, developed in our previous papers \cite{charmonium,Ferretti:2014xqa,bottomonium}.
This is a systematic way to perform coupled-channel calculations, by taking the effects of meson-meson intermediate states consistently into 
account at the quark level.
In our approach, the intermediate meson-meson states are weighted by means of $^3P_0$ amplitudes and the bare 
energies, entering the recursive coupled-channel equations for the physical masses of the mesons, are computed by using the numerically 
difficult Godfrey and Isgur's relativized QM \cite{Godfrey:1985xj}.
The UQM formalism will be extended to t-matrix calculations in a subsequent article.

In particular, in the present paper we intend to extend our previous calculation of the $c \bar c$ spectrum \cite{charmonium} to the 
$b \bar b$ one. 
Unlike the preliminary study of the lower bottomonium states of Ref. \cite{bottomonium}, in which the bare energies were taken as free 
parameters and their sum with the self energies fitted to the physical masses of the mesons of interest, thus losing predictive power, here 
we perform an explicit calculation of the bare energies within a potential model \cite{Godfrey:1985xj}.
Specifically, the new results we provide in this paper are: 1) the first systematic unquenching of a relativistic bottomonium quark model; 
2) the first discussion of the possibility of observing continuum effects in the $\chi_b(3P)$ system; and 3) the first systematic 
calculation of the open bottom strong decay widths of $b \bar b$ states within the $^3P_0$ model.

In Sec. \ref{Introduction}, we discuss the previous attempts made to unquench the quark model; in Sec. \ref{Formalism}, we recall the 
main ingredients of our UQM formalism and, in particular, in subsections \ref{An unquenched quark model for bottomonia} and 
\ref{3P0 pair-creation model} the main modifications we introduced in the $^3P_0$ operator (mainly the use of an effective $^3P_0$ 
strength that suppresses heavy quark pair-creation); in Sec. \ref{Results}, we show our results for the open bottom strong decay widths 
(subsection \ref{Open bottom strong decays}) and for the bottomonium spectrum with self energy corrections (subsection 
\ref{Bare and self energy calculation}) and finally discuss the nature of the $\chi_b(3P)$ system (subsection \ref{ChiB(3P)}).

\section{Formalism}
\label{Formalism} 
\subsection{Self energies}
The Hamiltonian we consider, 
\begin{equation}
	\label{eqn:Htot}
	H = H_0 + V  \mbox{ },
\end{equation}
is the sum of a first part, $H_0$, acting only in the bare meson space, and a second part, $V$, which can couple a meson state 
$\left| A \right\rangle$ to the meson-meson continuum $\left| BC \right\rangle$. 

The dispersive equation, resulting from a nonrelativistic Schr\"odinger equation, is
\begin{equation}
	\label{eqn:self-a}
	\Sigma(E_a) = \sum_{BC} \int_0^{\infty} q^2 dq \mbox{ } \frac{\left| V_{a,bc}(q) \right|^2}{E_a - E_{bc}}  \mbox{ },
\end{equation}
where the bare energy $E_a$ satisfies: 
\begin{equation}
	\label{eqn:self-trascendental}
	M_a = E_a + \Sigma(E_a)  \mbox{ }.
\end{equation}
$M_a$ is the physical mass of the meson $A$, with self energy $\Sigma(E_a)$. 
In Eq. (\ref{eqn:self-a}) one has to take the contributions from various meson-meson intermediate states $\left| BC \right\rangle$ into account. These channels, with relative momentum $q$ between $B$ and $C$, have quantum numbers $J_{bc}$ and $\ell$ coupled to the total angular momentum of the initial state $\left| A \right\rangle$. $V_{a,bc}$ stands for the coupling due to the operator $V$ between the intermediate state $\left| BC \right\rangle$ and the unperturbed quark-antiquark wave function of the meson $A$; $E_{bc} = E_b + E_c$ is the total energy of the channel $BC$, calculated in the rest frame.
Finally, if the bare energy of the meson $A$, $E_a$, is greater the threshold $E_{bc}$, the self energy of Eq. (\ref{eqn:self-a}) contains poles and is a complex number: in this case one has real loops instead of virtual ones. 

Since the physics of the dynamics depends on the matrix elements $V_{a,bc}(q)$, one has to choose a precise form for the transition 
operator, $V$, which is responsible for the creation of $q \bar q$ pairs. Our choice is that of the unquenched quark model (UQM) of Refs. 
\cite{bottomonium,charmonium}.

\subsection{An unquenched quark model for bottomonia}
\label{An unquenched quark model for bottomonia} 
In the unquenched quark model for mesons \cite{bottomonium,charmonium} the effects of quark-antiquark pairs are introduced explicitly into the quark model through a QCD-inspired $^{3}P_0$ pair-creation mechanism. 
This approach, which is a generalization of the unitarized quark model by T\"ornqvist and Zenczykowski \cite{Tornqvist} (see also Ref. \cite{Geiger:1996re}) is based on a QM, to which $q \bar q$ pairs with vacuum quantum numbers are added as a perturbation and where the pair-creation mechanism is inserted at the quark level. 

Under these assumptions, the meson wave function is made up of a zeroth order quark-antiquark configuration plus a sum over  the possible 
higher Fock components, due to the creation of $^{3}P_0$ $q \bar q$ pairs. Thus, one has  
\begin{eqnarray} 
	\label{eqn:Psi-A}
	\mid \psi_A \rangle &=& {\cal N} \left[ \mid A \rangle 
	+ \sum_{BC \ell J} \int d \vec{q} \, \mid BC \vec{q} \, \ell J \rangle \right.
	\nonumber\\
	&& \hspace{2cm} \left.  \frac{ \langle BC \vec{q} \, \ell J \mid T^{\dagger} \mid A \rangle } 
	{E_a - E_b - E_c} \right] ~, 
\end{eqnarray}
where $T^{\dagger}$ stands for the $^{3}P_0$ quark-antiquark pair-creation operator \cite{bottomonium,charmonium}, which depends on an 
effective pair-creation strength $\gamma_0^{\mbox{eff}}$, $A$ is the meson, $B$ and $C$ represent the intermediate state mesons, and $E_a$, 
$E_b = \sqrt{M_b^2 + q^2}$ and $E_c = \sqrt{M_c^2 + q^2}$ are the corresponding energies, $\vec{q}$ and $\ell$ the relative radial momentum 
and orbital angular momentum between $B$ and $C$ and $\vec{J} = \vec{J}_b + \vec{J}_c + \vec{\ell}$ is the total angular momentum. 
The wave functions of the mesons $A$, $B$ and $C$ can be written as harmonic oscillator wave functions, which depend on a single oscillator 
parameter $\alpha=0.5$ GeV. 

The $^{3}P_0$ quark-antiquark pair-creation operator, $T^{\dagger}$, is given by \cite{bottomonium,charmonium}
\begin{eqnarray}
\label{eqn:Tdag}
T^{\dagger} &=& -3 \, \gamma_0^{\mbox{eff}} \, \int d \vec{p}_3 \, d \vec{p}_4 \, 
\delta(\vec{p}_3 + \vec{p}_4) \, C_{34} \, F_{34} \,  
{e}^{-r_q^2 (\vec{p}_3 - \vec{p}_4)^2/6 }\, 
\nonumber\\
&& \hspace{0.5cm}  \left[ \chi_{34} \, \times \, {\cal Y}_{1}(\vec{p}_3 - \vec{p}_4) \right]^{(0)}_0 \, 
b_3^{\dagger}(\vec{p}_3) \, d_4^{\dagger}(\vec{p}_4) ~,   
\label{3p0}
\end{eqnarray}
where $b_3^{\dagger}(\vec{p}_3)$ and $d_4^{\dagger}(\vec{p}_4)$ are the creation operators for a quark and an antiquark with momenta 
$\vec{p}_3$ and $\vec{p}_4$, respectively. 
The $q \bar q$ pair is characterized by a color singlet wave function $C_{34}$, a flavor singlet wave function $F_{34}$, a spin triplet 
wave function $\chi_{34}$ with spin $S=1$ and a solid spherical harmonic ${\cal Y}_{1}(\vec{p}_3 - \vec{p}_4)$, which indicates that the 
quark and antiquark are in a relative $P$ wave. 
Since the operator $T^{\dagger}$ creates a pair of constituent quarks with an effective size, the pair-creation point has to be smeared out 
by a Gaussian factor, whose width $r_q$ has been determined from meson decays to be in the range $0.25-0.35$ fm 
\cite{Geiger:1996re,Geiger-Isgur,SilvestreBrac:1991pw}. 
In our calculation, we take the value $r_q = 0.335$ fm \cite{charmonium}.
The pair-creation strength, $\gamma_0^{\mbox{eff}}= \frac{m_n}{m_i} \mbox{ } \gamma_0$, is fitted to the strong decay $\Upsilon (4S)\rightarrow B\bar{B}$, and the value for $\gamma_0$ is extracted.  

In short, the two main differences from the old $^3P_0$ model are the introduction of a quark form factor, as already done by many authors 
such as T\"ornqvist and Zenczykowski \cite{Tornqvist}, Silvestre-Brac and Gignoux \cite{SilvestreBrac:1991pw} and Geiger and Isgur 
\cite{Geiger:1996re,Geiger-Isgur}, and the use of the effective strength $\gamma_0^{\mbox{eff}}= \frac{m_n}{m_i} \mbox{ } \gamma_0$, since 
it is well known that heavy flavor pair-creation is suppressed. 
We think that both these improvements, i.e. the introduction of the quark form factor and the effective strength $\gamma_0^{\mbox{eff}}$, 
already used in Refs. \cite{charmonium,bottomonium}, can make the model more realistic.

The matrix elements of the pair-creation operator $T^{\dagger}$ were derived in explicit form in the harmonic oscillator basis in Ref. 
\cite{Roberts:1992}, using standard Jacobi coordinates. 
\begin{figure}[htbp]
\begin{center}
\includegraphics[width=8cm]{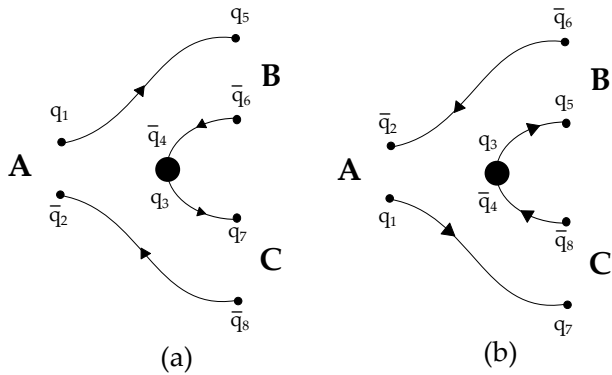}
\end{center}
\caption{Two diagrams can contribute to the process $A\rightarrow BC$. $q_{i}$ and $\bar{q}_{i}$ stand for the various initial ($i=1-4$) 
and final ($i=5-8$) quarks or antiquarks, respectively. Picture from Ref. \cite{bottomonium}; APS copyright.} 
\label{fig:diagrammi3P0}
\end{figure} 

In the UQM, the coupling $V_{a,bc}$ between the meson-meson continuum, $BC$, and the unperturbed wave function of the meson $A$ can be 
written as 
\begin{equation}
	\label{eqn:Vabc(q)}
	V_{a,bc}(q) = \sum_{\ell J} \left\langle BC \vec q  \, \ell J \right| T^\dag \left| A \right\rangle  \mbox{ }.
\end{equation}
In general, two different diagrams can contribute to the transition matrix element 
$\left\langle BC \vec q  \, \ell J \right| T^\dag \left| A \right\rangle $ (see Fig. \ref{fig:diagrammi3P0}): in the first one, the quark 
in $A$ ends up in $B$, while in the second one it ends up in $C$. In the majority of cases, one of these two diagrams vanishes; however, 
for some matrix elements, both must be taken into account \cite{bottomonium}, as, for example, in the case of the coupling 
$\eta_b \rightarrow \Upsilon\Upsilon$, where the initial $\left| b \bar b \right\rangle$ state is coupled to the final state 
$\left| b \bar b; b \bar b \right\rangle$ and the created pair is a $b \bar b$ one.   

Finally, by substituting Eq. (\ref{eqn:Vabc(q)}) into Eq. (\ref{eqn:self-a}), we have:
\begin{equation}
	\label{eqn:Sigma_A}
	\Sigma(E_a) = \sum_{BC \ell J} \int_0^{\infty} q^2 dq \mbox{ } 
	\frac{\left|\left\langle BC \vec q  \, \ell J \right| T^\dag \left| A \right\rangle \right|^2} {E_a - E_b - E_c} 
	\mbox{ }.
\end{equation}  

The values of the pair-creation model's parameters, used to compute the strong decays of Sec. \ref{Strong decay widths} and the vertices $\left\langle BC \vec q  \, \ell J \right| T^\dag \left| A \right\rangle$ of Eq. (\ref{eqn:Sigma_A}), are reported in Table \ref{tab:parameters}.
\begin{table}[htbp]  
\begin{center}
\begin{tabular}{cc} 
\hline 
\hline \\
Parameter  &  Value    \\ \\
\hline \\
$\gamma_0$ & 0.732     \\  
$\alpha$   & 0.500 GeV \\  
$r_q$      & 0.335 fm  \\
$m_n$      & 0.330 GeV \\
$m_s$      & 0.550 GeV \\
$m_c$      & 1.50 GeV  \\
$m_b$      & 4.70 GeV  \\   \\
\hline 
\hline
\end{tabular}
\end{center}
\caption{Pair-creation model parameters.}
\label{tab:parameters}  
\end{table}

\subsection{$^3P_0$ pair-creation model}
\label{3P0 pair-creation model}
In the $^3P_0$ pair-creation model \cite{3P0}, the open flavor strong decays of $b \bar b$ mesons take place via the production of a light 
$q \bar q$ pair (i.e. $q = u$, $d$ or $s$), with vacuum, i.e. $^3P_0$, quantum numbers, followed by the separation of the initial meson 
into two open-bottom mesons. 

The most recent variants of the $^3P_0$ model include a quark form factor in the transition operator 
\cite{charmonium,Geiger:1996re,Geiger-Isgur,Bijker:2009up,Santopinto:2010zza,Bijker:2012zza,bottomonium} that takes the non point-like 
nature of the constituent quarks into account, and an effective pair-production strength $\gamma_0^{\mbox{eff}}$ that suppresses 
unphysical heavy $q \bar q$ pair-creation \cite{charmonium,Kalashnikova:2005ui,bottomonium}. 

In particular, in Ref. \cite{Kalashnikova:2005ui} it is stated that, in the old $^3P_0$ model approach the pair-creation is 
flavor-independent, which implies an enhancement of the creation of heavy quarks in comparison with that of light quarks, without a 
fundamental reason for that. 
Thus, an effective pair-creation strength $\gamma_0^{\mbox{eff}}$ \cite{Kalashnikova:2005ui,charmonium,bottomonium}, defined as
\begin{equation}
	\label{eqn:gamma0-eff}
	\gamma_0^{\mbox{eff}} = \frac{m_n}{m_i} \mbox{ } \gamma_0  \mbox{ },
\end{equation}
is introduced, with $i$ = $n$ (i.e. $u$ or $d$), $s$, $c$ and $b$ (see Table \ref{tab:parameters}).
This problem has already been recognized and corrected by several authors \cite{Kalashnikova:2005ui,charmonium,bottomonium}.
The same mechanism as in Eq. (\ref{eqn:gamma0-eff}), including also a quark form factor, is used in the calculations of the present paper 
and of Refs. \cite{charmonium,bottomonium}.

\subsection{Godfrey and Isgur's relativized quark model}
\label{Godfrey and Isgur's relativized constituent quark model}
The relativized QM \cite{Godfrey:1985xj} is a potential model for $q \bar q$ meson spectroscopy, which was developed in 1985 by Godfrey and 
Isgur (see also Ref. \cite{Godfrey:2004ya}). 

The starting Hamiltonian of the model \cite{Godfrey:1985xj} is given by
\begin{equation}
	H = \sqrt{q^2 + m_1^2} + \sqrt{q^2 + m_2^2} + V_{\mbox{conf}} + V_{\mbox{hyp}} + V_{\mbox{so}}  \mbox{ },
	\label{eqn:H-GI}
\end{equation}
where $m_1$ and $m_2$ are the masses of the constituent quark and antiquark inside the meson, $q$ is their relative momentum (with 
conjugate coordinate $r$), and $V_{\mbox{conf}}$, $V_{\mbox{hyp}}$ and $V_{\mbox{so}}$ are the confining, hyperfine and spin-orbit potentials, 
respectively.

The confining potential \cite{Godfrey:1985xj},
\begin{equation}
	V_{\mbox{conf}} = - \left(\frac{3}{4} \mbox{ } c + \frac{3}{4} \mbox{ } br -\frac{\alpha_s(r)}{r} \right) 
	\vec F_1 \cdot \vec F_2  \mbox{ },
\end{equation}
contains a constant, $c$, a linear confining term and a Coulomb-like interaction, which depends on a QCD-motivated running coupling 
constant $\alpha_s(r)$.

The hyperfine interaction is written as \cite{Godfrey:1985xj}
\begin{equation}
	\label{eqn:Vhyp}
	\begin{array}{rcl}
	V_{\mbox{hyp}} & = & -\frac{\alpha_s(r)}{m_{1}m_{2}} \left[\frac{8\pi}{3} \vec S_{1} \cdot \vec S_{2} \mbox{ }
	\delta ^{3}(\vec r) \right. \\ 
	& + & \left. \frac{1}{r^{3}} \left( \frac{3 \mbox{ } \vec S_{1} \cdot \vec r \mbox{ } 
	\vec S_{2} \cdot \vec r}{r^{2}} - \vec S_{1} \cdot \vec S_{2}\right) \right] \mbox{ } 
	\vec F_{i} \cdot \vec F_{j}  \mbox{ }.
	\end{array}
\end{equation}

The spin-orbit potential \cite{Godfrey:1985xj}, 
\begin{equation}
	V_{\mbox{so}} = V_{\mbox{so,cm}} + V_{\mbox{so,tp}}  \mbox{ },
\end{equation}
is the sum of two contributions, where
\begin{subequations}
\begin{equation}
	\begin{array}{rcl}
	V_{\mbox{so,cm}} & = & - \frac{\alpha_s(r)}{r^{3}} \left( \frac{1}{m_{i}}+\frac{1}{m_{j}} \right) \\
	& & \left( \frac{\vec S_{i}}{m_{i}}+\frac{\vec S_{j}}{m_{j}} \right) \cdot \vec L 
	\;\;\vec F_{i}\cdot \vec F_{j}
	\end{array}
\end{equation}
is the color-magnetic term and
\begin{equation}
	V_{\mbox{so,tp}} = - \frac{1}{2r}\frac{\partial H_{ij}^{conf}}{\partial r} \left( \frac{\vec S_{i}} 
	{m_{i}^{2}}+\frac{\vec S_{j}}{m_{j}^{2}}\right) \cdot \vec L
\end{equation}
\end{subequations}
is the Thomas-precession term. 

What is known as Godfrey and Isgur's model \cite{Godfrey:1985xj} is the Hamiltonian of Eq. (\ref{eqn:H-GI}) plus some relativistic effects. 
These effects include the introduction of a potential smearing and the replacement of factors of quark mass by quark kinetic energy. 
This is exactly the model used in the present paper for the bare energy calculation of Sec. \ref{Bare and self energy calculation}.

\section{Results}
\label{Results} 
\subsection{Open-bottom strong decays in the $^3P_0$ pair-creation model}
\label{Open bottom strong decays} 
\label{Strong decay widths} 
In this section, we show our calculation of the open-bottom strong decay widths of higher bottomonia (see table \ref{tab:strong-decays}).
The decay widths are calculated within the $^3P_0$ model \cite{charmonium,bottomonium} as
\begin{equation}
\label{eqn:3P0width}
	\Gamma_{A \rightarrow BC} = \Phi_{A \rightarrow BC}(q_0) \sum_{\ell, J} 
	\left| \left\langle BC \vec q_0  \, \ell J \right| T^\dag \left| A \right\rangle \right|^2 \mbox{ }.
\end{equation}
Here, $\Phi_{A \rightarrow BC}(q_0)$ is the standard relativistic phase space factor \cite{charmonium,bottomonium}, 
\begin{equation}
	\label{eqn:relPSF}    
	\Phi_{A \rightarrow BC} = 2 \pi q_0 \frac{E_b(q_0) E_c(q_0)}{M_a}  \mbox{ },
\end{equation}
which depends on the relative momentum $q_0$ between $B$ and $C$ and on the energies of the two intermediate state mesons, 
$E_b = \sqrt{M_b^2 + q_0^2}$ and $E_c = \sqrt{M_c^2 + q_0^2}$. 
The values of the masses $M_a$, $M_b$ and $M_c$ used in the calculation are taken from the PDG \cite{Nakamura:2010zzi} and Ref. 
\cite{Mizuk:2012pb}, while in the case of still unobserved states we use Godfrey and Isgur's model predictions, obtained with the values of 
the model's parameters shown in Table \ref{tab:Parametri-Godfrey.Isgur} (see Tables \ref{tab:open-bottom} and \ref{tab:strong-decays}, 
second column).
In our calculation, we use a variational basis of 200 harmonic oscillator shells, so that our results converge very well.

\begin{table}[htbp]  
\begin{center}
\begin{tabular}{cccc}
\hline
\hline \\
State   & $J^P$ & $M$ [GeV] & Source \\  \\
$B$     & $0^-$ & 5.279     & \cite{Nakamura:2010zzi} \\   
$B^*$   & $1^-$ & 5.325     & \cite{Nakamura:2010zzi} \\  
$B_s$   & $0^-$ & 5.366     & \cite{Nakamura:2010zzi} \\  
$B_s^*$ & $1^-$ & 5.416     & \cite{Nakamura:2010zzi} \\  
$B_c$   & $0^-$ & 6.277     & \cite{Nakamura:2010zzi} \\  
$B_c^*$ & $1^-$ & 6.340     & \cite{Godfrey:1985xj}   \\  \\
\hline
\hline
\end{tabular}
\end{center}
\caption{Masses of open-bottom mesons used in the calculations. When available, we used the experimental values form PDG 
\cite{Nakamura:2010zzi}; otherwise, in the case of the $B_c^*$ resonance, we took the theoretical prediction of the relativized QM 
\cite{Godfrey:1985xj}.}
\label{tab:open-bottom}
\end{table}

The operator $T^\dag$ inside the $^{3}P_0$ amplitudes $\left\langle BC	\vec q_0  \, \ell J \right| T^\dag \left| A \right\rangle$ is that 
of Eq. (\ref{eqn:Tdag}), which also contains the quark form factor of Refs. \cite{Geiger:1996re,Geiger-Isgur}.
The quark form factor, which takes the non point-like nature of the constituent quarks into account, is not included in the original 
formulation of the $^3P_0$ model \cite{3P0}. 
Another difference between our calculation and those of Refs. \cite{3P0,Roberts:1992,Ackleh:1996yt,Barnes:2005pb} is the substitution of 
the pair-creation strength $\gamma_0$ by the effective strength $\gamma_0^{\mbox{eff}}$ of Eq. (\ref{eqn:gamma0-eff}). 
The introduction of this effective mechanism suppresses those diagrams in which a heavy $q \bar q$ pair is created, as discussed in Sec. 
\ref{3P0 pair-creation model}. 
More details on this mechanism can be found in Refs. \cite{bottomonium,charmonium,Kalashnikova:2005ui}.

Finally, the results of our calculation, obtained with the values of the model parameters shown in Table \ref{tab:parameters}, are reported 
in Table \ref{tab:strong-decays}. See also Table \ref{tab:open-bottom-exp}, where our theoretical results are compared with the existing 
experimental data \cite{Nakamura:2010zzi}. 
Here, we also provide a rough evaluation of the theoretical error on the predicted decay widths.
The error is estimated considering the upper and lower limits on the values of the experimental masses \cite{Nakamura:2010zzi}.
For example, in the case of the $\Upsilon(4^3S_1)$,
\begin{equation}
	M_{\Upsilon(4^3S_1)} = 10579.4 \pm 1.2 \mbox{ MeV} \mbox{ },
\end{equation} 
we have $M_{\Upsilon(4^3S_1)}^{min} = 10578.2$ MeV and $M_{\Upsilon(4^3S_1)}^{max} = 10580.6$ MeV. Analogously, in the case of $B$, we have $M_B^{min} = 5279.09$ MeV and $M_B^{max} = 5279.43$ MeV. Different combinations between the upper and lower limits of the masses of the decaying and final state mesons provide minimum and maximum values for the theoretical decay width, from which one can extract the theoretical error.

\begin{table*}
\begin{tabular}{ccccccccc} 
\hline 
\hline \\
Meson                            &  Mass [MeV] &  $J^{PC}$    & $B \bar B$ & $B \bar B^*$ & $B^* \bar B^*$ & $B_s \bar B_s$ & $B_s \bar B_s^*$ & $B_s^* \bar B_s^*$ \\ 
                 &       &       &           & $\bar B B^*$ &                         &                         & $\bar B_s B_s^*$                                 \\ \\
\hline \\
$\Upsilon(10580)$ or $\Upsilon(4^3S_1)$      & 10.595                            & $1^{--}$   & 20 & -- & -- &  --   &  --  &  --  \\
                                    & $10579.4\pm1.2^\dag$ &                  &       &     &     &         &        &        \\
$\chi_{b2}(2^3F_2)$     & $10585$ & $2^{++}$ & 34  & -- & -- &  --  &  --  &  --  \\
$\Upsilon(3^3D_1)$     & $10661$ & $1^{--}$   & 23 &  4  & 15 &  --  &  --  &  --  \\
$\Upsilon_2(3^3D_2)$ & $10667$ & $2^{--}$ &   -- & 37  & 30 &  --  &  --  &  --  \\
$\Upsilon_2(3^1D_2)$ & $10668$ & $2^{-+}$ &  --  & 55  & 57 &  --  &  --  &  --  \\  
$\Upsilon_3(3^3D_3)$ & $10673$ & $3^{--}$ &  15  & 56  &113 &  -- &  --  &  --  \\ 
$\chi_{b0}(4^3P_0)$     & $10726$ & $0^{++}$ & 26  & -- & 24 &  --  &  --  & --  \\
$\Upsilon_3(2^3G_3)$ & $10727$ & $3^{--}$ &   3   &  43  & 39 &  -- &  --  &  -- \\
$\chi_{b1}(4^3P_1)$     & $10740$ & $1^{++}$ & --  & 20 &  1  &  --  &  --  &  --  \\
$h_b(4^1P_1)$             & $10744$ & $1^{+-}$ &  --  & 33 &  5  &  --  &  --  &  --  \\
$\chi_{b2}(4^3P_2)$     & $10751$ & $2^{++}$ & 10  & 28 &  5  &   1  &  --  &  --  \\
$\chi_{b2}(3^3F_2)$     & $10800$ & $2^{++}$ &  5   & 26 & 53 &   2  &   2   &  --  \\
$\Upsilon_3(3^1F_3)$  & $10803$ & $3^{+-}$ &  --  & 28   & 46  &  --  &  3 &  --  \\
$\Upsilon(10860)$ or $\Upsilon(5^3S_1)$      & $10876\pm11^\dag$ & $1^{--}$  &  1  & 21 & 45 &   0   &   3   &   1  \\
$\Upsilon_2(4^3D_2)$ & $10876$ & $2^{--}$ &  --  & 28  & 36 &  --  &   4   &   4   \\
$\Upsilon_2(4^1D_2)$ & $10877$ & $2^{-+}$ &  --  & 22  & 37 &  --  &   4   &   3   \\
$\Upsilon_3(4^3D_3)$ & $10881$ & $3^{--}$ &   1   &  4    & 49 &   0  &   1  &   2  \\
$\Upsilon_3(3^3G_3)$ & $10926$ & $3^{--}$ &   7   &  0    & 13 &   2   &   0  &  5   \\
$\Upsilon(11020)$ or $\Upsilon(6^3S_1)$      & $11019\pm8^\dag$ & $1^{--}$  &  0  &  8  & 26 &   0   &   0   &   2  \\   \\
\hline 
\hline
\end{tabular}
\caption{Strong decay widths (in MeV) in heavy meson pairs for higher bottomonium states. Column 2 shows the values of the masses of the decaying $b \bar b$ states: when available, we used the experimental values from PDG \cite{Nakamura:2010zzi} ($\dag$), otherwise the theoretical predictions of the relativized QM \cite{Godfrey:1985xj}, whose mass formula we have re-fitted to the most recent experimental data (parameters as from Table \ref{tab:Parametri-Godfrey.Isgur}). Columns 3-8 show the decay width contributions from various $BC$ channels, such as $B \bar B$, $B \bar B^*$ and so on. The values of the $^3P_0$ model parameters, fitted to experimental data for the strong decay widths of $b \bar b$ resonances (see App. \ref{Model parameters}), are shown in Table \ref{tab:parameters}. The symbol -- in the table means that a certain decay is forbidden by selection rules or that the decay cannot take place because it is below the threshold.} 
\label{tab:strong-decays}  
\end{table*}

\begin{table}[htbp]  
\begin{center}
\begin{tabular}{llllll}
\hline
\hline \\
$m_b$                              & = 5.024 GeV &$b$                        & = 0.156 GeV$^2$   & $\alpha_s^{\mathrm{cr}}$ & = 0.60          \\
$\Lambda$                       & = 0.200 GeV & $c$                       & = $-$0.280 GeV     & $\sigma_0$                       & = 0.146 GeV \\ 
$s$                                   & = 4.36          & $\epsilon_c$         & = $-$0.242            & $\epsilon_t$                      & = 0.030         \\ 
$\epsilon_{so(V)}$            & = $-$0.053  & $\epsilon_{so(S)}$  & = 0.019                  & &      \\ \\
\hline
\hline
\end{tabular}
\end{center}
\caption{Resulting values of Godfrey and Isgur's model \cite{Godfrey:1985xj} parameters, obtained by re-fitting the mass formula of Eq. 
(\ref{eqn:H-GI}) with the most recent experimental data \cite{Nakamura:2010zzi}.}
\label{tab:Parametri-Godfrey.Isgur}
\end{table}

\begin{table}[htbp]  
\begin{center}
\begin{tabular}{ccc}
\hline
\hline \\
State              & $\Gamma_{\mbox{theor}}$ $(^3P_0)$ [MeV] & $\Gamma_{\mbox{exp}}$ [MeV] \\  \\
\hline \\
$\Upsilon(4^3S_1)$ & $20.49^{+0.01}_{-0.12}$                           & $20.5\pm2.5$  \\ 
$\Upsilon(10860)$  & $71.08^{+24.34}_{-13.83}$                        & $42^{+29}_{-24}$  \\    \\
\hline
\hline
\end{tabular}
\end{center}
\caption{Our results for the open-bottom strong decay widths of Table \ref{tab:strong-decays} are compared to the existing experimental data \cite{Nakamura:2010zzi}. We also provide an estimation of the theoretical error, which is determined considering the upper and lower limits on the values of the experimental masses from PDG \cite{Nakamura:2010zzi}. Different combinations between the upper and lower limits of the masses of the decaying and final state mesons provide minimum and maximum values for the theoretical decay width, from which one can extract the theoretical error.}
\label{tab:open-bottom-exp}
\end{table}

To obtain results for the masses of the higher lying $b \bar b$ resonances, we use the relativized QM of Ref. \cite{Godfrey:1985xj}, whose 
mass formula we have re-fitted to the most recent experimental data (see Table \ref{tab:Parametri-Godfrey.Isgur}). 
Something similar was done in Ref. \cite{Barnes:2005pb} for charmonia. 

This re-fit was necessary to compute the strong decays, which require precise values for the masses of the decaying mesons, including the 
higher-lying states. Indeed, Godfrey and Isgur's 85 original results \cite{Godfrey:1985xj} show a deviation from the most recent 
experimental data of the order of 50 MeV in the case of $4S$ states. Godfrey and Isgur's prediction for the mass of $\Upsilon(4S)$ 
(10.63 GeV \cite{Godfrey:1985xj}) is approximately 50 MeV higher than the corresponding experimental data ($10579.4\pm1.2$ MeV 
\cite{Nakamura:2010zzi}), moreover, their theoretical prediction for the mass of $\eta_b(4S)$ (10.62 GeV) is 40 MeV higher than the mass of 
$\Upsilon(4S)$, while on contrary an $\eta_b(4S)$ state should be lower in energy. 
The value of $\eta_b(4S)$'s mass, which was absent in the original paper of 1985 \cite{Godfrey:1985xj}, was extracted by running a 
numerical program that calculates Godfrey and Isgur model's spectrum with the original values of the parameters as reported in Ref. 
\cite{Godfrey:1985xj}.
The $4S$ resonances are important, being the lowest energy $b \bar b$ states that decay into two open-bottom mesons. 
Since we are interested in calculating observables (the strong decay widths) that have a strong dependence on the masses of the mesons 
involved in the calculation, we thought that it was important to update Godfrey and Isgur's 1985 results in the $b \bar b$ sector. 
At that time, many $b \bar b$ states were still unobserved. Moreover, since Godfrey and Isgur's results differ from the experimental data 
in the $4S$ case, we think that this might also be the case of other higher lying radial excitations, such as $4P$. 
Thus, in our fit, we preferred to get a better reproduction of the radial excitations instead of the low-lying ones, because the latter are 
useless in computing the decays.

\subsection{Bare and self energy calculation of $b \bar b$ states}
\label{Bare and self energy calculation}
The relativized QM \cite{Godfrey:1985xj} is now used to compute the bare energies of the $b \bar b$ mesons, $E_a$'s, at each step of an 
iterative procedure. 
Indeed, in this case, the quantities fitted to the spectrum of bottomonia \cite{Nakamura:2010zzi,Mizuk:2012pb} are the physical masses 
$M_a$'s of Eq. (\ref{eqn:self-trascendental}) and therefore the fitting procedure is an iterative one.

Indeed, once the values of the bare energies are known, it is possible to calculate the self energies $\Sigma(E_a)$'s of the $b \bar b$ 
states through Eq. (\ref{eqn:Sigma_A}), summing over a complete set of accessible SU$_{\mbox{f}}$(5)$\otimes$SU$_{\mbox{spin}}$(2) $1S$ 
intermediate states.
If the bare energy of the initial meson $A$ is above the threshold $BC$, i.e. $E_a > M_b + M_c$, the self energy contribution due to the 
meson-meson $BC$ channel is computed as
\begin{equation}
	\label{eqn:SigmaE.real.loops}
	\begin{array}{l}
	\Sigma(E_a)(BC) = \\
	\hspace{0.5cm} \mathcal{P} \int_{M_b+M_c}^{\infty} \frac{dE_{bc}}{E_a - E_{bc}} \mbox{ } 
	\frac{q E_b E_c}{E_{bc}} \left| \left\langle BC \vec q \, \ell J \right| T^\dag \left| A \right\rangle \right|^2 \\
	\hspace{0.5cm} + \mbox{ } 2 \pi i \left\{ \frac{q E_b E_c}{E_a} 
	\left| \left\langle BC \vec q \, \ell J \right| T^\dag \left| A \right\rangle \right|^2 \right\}_{E_{bc} = E_a} 
	\mbox{ },
	\end{array}
\end{equation}
where the symbol $\mathcal{P}$ indicates a principal part integral, calculated numerically, and 
$2 \pi i \left\{ \frac{q E_b E_c}{E_a} \left| \left\langle BC \vec q \, \ell J \right| T^\dag \left| A \right\rangle \right|^2 \right\}_{E_{bc} = E_a}$ 
is the imaginary part of the self energy.

\begin{table}[htbp]  
\begin{center}
\begin{tabular}{llllll}
\hline
\hline \\
$m_b$                   & = 4.568 GeV & $b$                       & = 0.1986 GeV$^2$ & $\alpha_s^{\mathrm{cr}}$ & = 0.600    \\
$\Lambda$            & = 0.200 GeV & $c$                       & = 0.628 GeV     & $\sigma_0$                       & = 0.0127 GeV \\ 
$s$                        & = 2.655        & $\epsilon_c$         & = $-$0.2948          & $\epsilon_t$                      & = 0.0129   \\ 
$\epsilon_{so(V)}$ & = $-$0.0715 & $\epsilon_{so(S)}$ & = 0.0573                &                                          &            \\ \\
\hline
\hline
\end{tabular}
\end{center}
\caption{Values of Godfrey and Isgur's model parameters, obtained by fitting the results of Eq. (\ref{eqn:self-trascendental}) to the 
experimental data \cite{Mizuk:2012pb,Nakamura:2010zzi}.}
\label{tab:Parametri-self}
\end{table}

\begin{table*}
\begin{tabular}{cccccccccccccccccc} 
\hline 
\hline \\
State               & $J^{PC}$ & $B \bar B$ & $B\bar B^*$ & $B^*\bar B^*$ & $B_s \bar B_s$ & $B_s \bar B_s^*$ & $B_s^* \bar B_s^*$ & $B_c \bar B_c$ & $B_c \bar B_c^*$ & $B_c^* \bar B_c^*$ & $\eta_b \eta_b$ & $\eta_b \Upsilon$ & $\Upsilon \Upsilon$ & $\Sigma(E_a)$ & $E_a$ & $M_a$ & $M_{exp.}$ \\
                    &          &            & $\bar BB^*$ &               &                & $\bar B_s B_s^*$ &                     &                & $\bar B_c B_c^*$ &                    &                 &                   &                      &               &       &       &            \\ \\
\hline \\
$\eta_b(1^1S_0)$  & $0^{-+}$ &   --  &  -26  & -26  & --  & -5  & -5  & -- &  -1 &  -1 & --  & --  &  0  & -64  & 9455 &  9391 & 9391 \\
$\Upsilon(1^3S_1)$  & $1^{--}$ &   -5  &  -19  & -32  & -1  & -4  & -7  & 0  &  0  &  -1 & --  & 0   &  -- & -69  & 9558 &  9489 & 9460 \\
$\eta_b(2^1S_0)$    & $0^{-+}$ &   --  &  -43  & -41  & --  & -8  & -7  & -- &  -1 &  -1 & --  & --  &  0  & -101 & 10081 & 9980 & 9999 \\
$\Upsilon(2^3S_1)$  & $1^{--}$ &   -8  &  -31  & -51  & -2  & -6  & -9  &  0 &  0  &  -1 & --  &  0  &  -- & -108 & 10130 &  10022 & 10023 \\
$\eta_b(3^1S_0)$    & $0^{-+}$ &   --  &  -59  & -52  & --  & -8  & -8  & -- &  -1 &  -1 & --  & --  &  0  & -129 & 10467 &  10338 & -- \\
$\Upsilon(3^3S_1)$  & $1^{--}$ &  -14  &  -45  & -68  & -2  & -6  & -10 &  0 & 0   &  -1 & --  &  0  &  -- & -146 & 10504 &  10358 & 10355 \\
$h_b(1^1P_1)$       & $1^{+-}$ &  --   &  -49  & -47  & --  & -9  & -8  & -- &  -1 &  -1 & --  &  0  &  -- & -115 & 10000 & 9885 & 9899 \\
$\chi_{b0}(1^3P_0)$ & $0^{++}$ &  -22  &   --  & -69  & -3  & --  & -13 &  0 &  -- &  -1 &  0  & --  &   0 & -108 & 9957 & 9849 & 9859 \\
$\chi_{b1}(1^3P_1)$ & $1^{++}$ &   --  &  -46  & -49  & --  & -8  & -9  & -- &  -1 &  -1 & --  & --  &  0  & -114 & 9993 & 9879 & 9893 \\
$\chi_{b2}(1^3P_2)$ & $2^{++}$ &  -11  &  -32  & -55  & -2  & -6  & -9  &  0 &  -1 &  -1 &  0  & --  &   0 & -117 & 10017 & 9900 & 9912 \\
$h_b(2^1P_1)$       & $1^{+-}$ &   --  &  -66  & -59  & --  & -10 & -9  & -- &  -1 &  -1 & --  &  0  &  -- & -146 & 10393 & 10247 & 10260 \\
$\chi_{b0}(2^3P_0)$ & $0^{++}$ &  -33  &   --  & -85  & -4  & --  & -14 &  0 &  -- &  -1 &  0  & --  &   0 & -137 & 10363 & 10226 & 10233 \\
$\chi_{b1}(2^3P_1)$ & $1^{++}$ &   --  &  -63  & -60  & --  & -9  & -10 & -- &  -1 &  -1 & --  & --  &  0  & -144 & 10388 & 10244 & 10255 \\
$\chi_{b2}(2^3P_2)$ & $2^{++}$ &  -16  &  -42  & -72  & -2  & -6  & -10 &  0 &   0 &  -1 &  0  & --  &   0 & -149 & 10406 & 10257 & 10269 \\
$h_b(3^1P_1)$       & $1^{+-}$ &   --  &  -18  & -73  & --  & -11 & -10 & -- &  -1 &  -1 & --  &  0  &  -- & -114 & 10705 & 10591 & --    \\
$\chi_{b0}(3^3P_0)$ & $0^{++}$ &  -4   &   --  & -160 & -6  & --  & -15 &  0 &  -- &  -1 &  0  & --  &   0 & -186 & 10681 & 10495 & --    \\
$\chi_{b1}(3^3P_1)$ & $1^{++}$ &   --  &  -25  & -74  & --  & -11 & -10 & --  &  0 &  -1 & --  & --  &  0  & -121 & 10701 & 10580 & --    \\
$\chi_{b2}(3^3P_2)$ & $2^{++}$ &  -19  &  -16  & -79  & -3  & -8  & -12 &  0  &   0 & -1 &  0  & --  &   0 & -138 & 10716 & 10578 & --    \\
$\Upsilon_2(1^1D_2)$  & $2^{-+}$ &   --  &  -72  & -66 & --  & -11 & -10 & -- &  -1 &  -1 & --  & --  &   0 & -161 & 10283 & 10122 & -- \\ 
$\Upsilon(1^3D_1)$  & $1^{--}$ &  -24  &  -22  & -90 & -3  & -3  & -16 &  0 &   0 &  -1 & --  &  0  &  -- & -159 & 10271 & 10112 & -- \\ 
$\Upsilon_2(1^3D_2)$  & $2^{--}$ &   --  &  -70  & -68 & --  & -10 & -11 & -- &  -1 &  -1 & --  &  0  &  -- & -161 & 10282 & 10121 & 10164 \\
$\Upsilon_3(1^3D_3)$  & $3^{--}$ &  -18  &  -43  & -78 & -3  & -8  & -11 &  0 &  -1 &  -1 & --  &  0  &  -- & -163 & 10290 & 10127 & -- \\ \\ 
\hline 
\hline
\end{tabular}
\caption{Self energies, $\Sigma(E_a)$ (in MeV, see column 15), for $1S$, $2S$, $3S$, $1P$, $2P$, $3P$ and $1D$ bottomonium states due to 
coupling to the meson-meson continuum, calculated with the effective pair-creation strength of Eq. (\ref{eqn:gamma0-eff}) and the values 
of the UQM parameters of Table \ref{tab:parameters}. Columns 3-14 show the contributions to $\Sigma(E_a)$ from various channels $BC$, such 
as $B \bar B$, $B \bar B^*$ and so on. In column 16 are reported the values of the bare energies, $E_a$, calculated within the relativized 
QM \cite{Godfrey:1985xj}, with the values of the model parameters of Table \ref{tab:Parametri-self}. In column 17 are reported the 
theoretical estimations $M_a$ of the masses of the $b \bar b$ states, which are the sum of the self energies $\Sigma(E_a)$ and the bare 
energies $E_a$ (see also Fig. \ref{fig:bottomonium-spectrum-self}). Finally, in column 18 are reported the experimental values of the 
masses of the $b \bar b$ states \cite{Mizuk:2012pb,Nakamura:2010zzi}. The symbol $-$ means that the contribution from a channel is 
suppressed by selection rules (spins, $G$-parity, ...).} 
\label{tab:Mass-shifts} 
\end{table*}

\begin{figure}[htbp]
\begin{center}
\includegraphics[width=7cm]{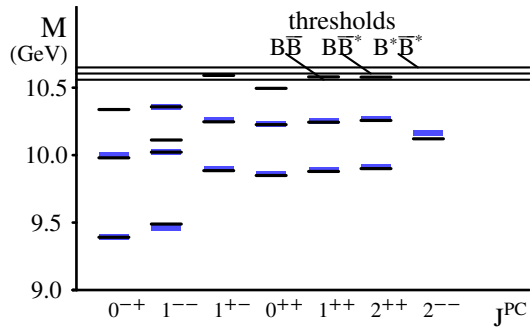}
\end{center}
\caption{Comparison between the calculated masses (black lines) of $1S$, $2S$, $3S$, $1P$, $2P$, $3P$ and $1D$ bottomonium states via Eq. (\ref{eqn:self-trascendental}) and the experimental ones \cite{Mizuk:2012pb,Nakamura:2010zzi} (boxes). The new values of Godfrey and Isgur's model parameters are taken from Table \ref{tab:Parametri-self}. The picture also shows the lowest strong decay thresholds.} 
\label{fig:bottomonium-spectrum-self}
\end{figure}

Finally, the results of our calculation, obtained with the set of parameters of Tables \ref{tab:parameters} and \ref{tab:Parametri-self} 
and the effective pair-creation strength of Eq. (\ref{eqn:gamma0-eff}), are given in Table \ref{tab:Mass-shifts} and Fig. \ref{fig:bottomonium-spectrum-self}. 
This means that the vertices $\left\langle BC \vec q  \, \ell J \right| T^\dag \left| A \right\rangle$ of Eqs. (\ref{eqn:Sigma_A}) and 
(\ref{eqn:SigmaE.real.loops}) are computed with the same set of $^3P_0$ model parameters as in Sec. \ref{Strong decay widths}, fitted to 
the experimental $\Upsilon(4S) \rightarrow B \bar B$ strong decay width \cite{Nakamura:2010zzi} (see App. \ref{Model parameters}). 

\subsection{$\chi_b(3P)$ system}
\label{ChiB(3P)} 
\begin{figure}[htbp]
\begin{center}
\includegraphics[width=8cm]{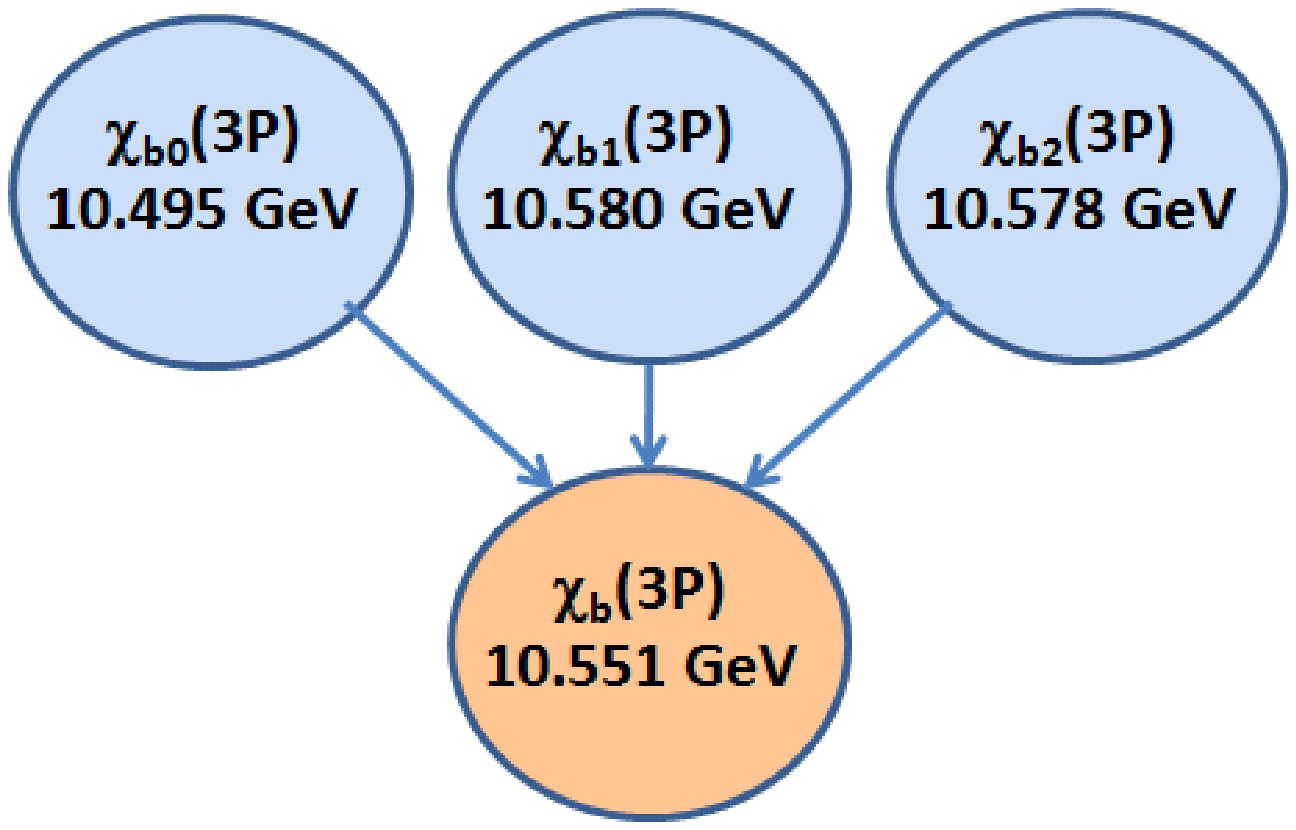}
\end{center}
\caption{Mass barycenter (in GeV) of the $\chi_b(3P)$ system in our UQM calculation.} 
\label{fig:chib3P}
\end{figure}

The $\chi_b(3P)$ system was discovered by the ATLAS Collaboration in 2012 \cite{Aad:2011ih} and then confirmed by the D0 Collaboration \cite{Abazov:2012gh}. 
Since the $\chi_b(3P)$ resonances lie quite close to $B \bar B$, $B \bar B^*$ and $B^* \bar B^*$ decay thresholds, their wave functions may contain important continuum components, as we have shown in the $c \bar c$ sector in the case of the $X(3872)$ \cite{charmonium}.
We think that the present experimental data \cite{Aad:2011ih,Abazov:2012gh} cannot exclude this possibility. 
In particular, in Ref. \cite{Abazov:2012gh} the authors state \begin{em}"Further analysis is underway to determine whether this structure is due to the $\chi_b(3P)$ system or some exotic bottom-quark state".\end{em}
Thus, we think that our results for these states, and in particular for the splittings between them, can be used to discuss this particular problem (see Fig. \ref{fig:chib3P} and Tables \ref{tab:mass-bary}, \ref{tab:mass-bary-GI} and \ref{tab:mass-bary-GI-OLD}). 
Indeed, the magnitude of the splittings between $\chi_b(3P)$ states of the multiplet is still unknown (see Ref. \cite{Dib:2012vw}). 

\begin{table}[htbp]  
\begin{center}  
\begin{tabular}{ccc}
\hline
\hline \\
$M^{\mbox{th}}_{\chi_b(1P)}$ &  $\Delta M_{21}(1P)$ & $\Delta M_{10}(1P)$    \\
9876                                       &  21                            & 30                               \\ \\
$M^{\mbox{th}}_{\chi_b(2P)}$ &  $\Delta M_{21}(2P)$ & $\Delta M_{10}(2P)$    \\
10242                                     &  13                            & 18                               \\ \\
$M^{\mbox{th}}_{\chi_b(3P)}$ &  $\Delta M_{21}(3P)$ & $\Delta M_{10}(3P)$    \\
10551                                     &  -2                            & 85                              \\ \\
\hline
\hline
\end{tabular}
\end{center}
\caption{Mass barycenters of $\chi_b(1P)$, $\chi_b(2P)$ and $\chi_b(3P)$ systems (column 1) and mass splittings between the members of the 
$\chi_b(1P)$, $\chi_b(2P)$ and $\chi_b(3P)$ multiplets (column 2 and 3), from Table \ref{tab:Mass-shifts}. These are the results of our UQM 
calculation of the $b \bar b$ spectrum with self energy corrections of Table \ref{tab:Mass-shifts}. The results are expressed in MeV. 
The notation $\Delta M_{21}(1P)$ stands for the mass difference between the $\chi_{b2}(1P)$ and $\chi_{b1}(1P)$ resonances, 
$\Delta M_{10}(1P)$ for the mass difference between the $\chi_{b1}(1P)$ and $\chi_{b0}(1P)$ resonances, and so on.}
\label{tab:mass-bary}
\end{table}
\begin{table}[htbp]  
\begin{center}
\begin{tabular}{ccc}
\hline
\hline \\
$M_{\chi_b(1P)}$ & $\Delta M_{21}(1P)$ & $\Delta M_{10}(1P)$    \\
9894                   & 21                            & 30                               \\ \\
$M_{\chi_b(2P)}$ & $\Delta M_{21}(2P)$ & $\Delta M_{10}(2P)$    \\
10241                 & 21                            & 14                               \\ \\
$M_{\chi_b(3P)}$ & $\Delta M_{21}(1P)$ & $\Delta M_{10}(1P)$    \\
10510                 & 17                            & 13                               \\ \\
\hline
\hline
\end{tabular}
\end{center}
\caption{Mass barycenters of $\chi_b(1P)$, $\chi_b(2P)$ and $\chi_b(3P)$ systems (column 1) and mass splittings between the members of the 
$\chi_b(1P)$, $\chi_b(2P)$ and $\chi_b(3P)$ multiplets (column 2 and 3), from Table \ref{tab:strong-decays}. These are the results of our 
re-fit of Godfrey and Isgur's mass formula, with the model parameters of Table \ref{tab:Parametri-Godfrey.Isgur}. 
The results are expressed in MeV.}
\label{tab:mass-bary-GI}
\end{table}
\begin{table}[htbp]  
\begin{center}
\begin{tabular}{ccc}
\hline
\hline \\
$M_{\chi_b(1P)}$ & $\Delta M_{21}(1P)$ & $\Delta M_{10}(1P)$    \\
9872                   & 21                            & 30                               \\ \\
$M_{\chi_b(2P)}$ & $\Delta M_{21}(2P)$ & $\Delta M_{10}(2P)$    \\
10244                 & 15                            & 21                               \\ \\
$M_{\chi_b(3P)}$ & $\Delta M_{21}(1P)$ & $\Delta M_{10}(1P)$    \\
10536                 & 12                            & 16                               \\ \\
\hline
\hline
\end{tabular}
\end{center}
\caption{Mass barycenters of $\chi_b(1P)$, $\chi_b(2P)$ and $\chi_b(3P)$ systems (column 1) and mass splittings between the members of the 
$\chi_b(1P)$, $\chi_b(2P)$ and $\chi_b(3P)$ multiplets (columns 2 and 3) within the relativized QM \cite{Godfrey:1985xj}. These results were 
extracted by running a numerical program that calculates Godfrey and Isgur model's spectrum for $b \bar b$ states with the values of the 
model parameters reported in the original paper of 1985 \cite{Godfrey:1985xj}. The results are expressed in MeV.}
\label{tab:mass-bary-GI-OLD}
\end{table}
As shown by Table \ref{tab:mass-bary}, our UQM result for the mass barycenter of the $\chi_b(3P)$ system [obtained by averaging the 
theoretical values of the masses of the $\chi_{b0}(3P)$, $\chi_{b1}(3P)$ and $\chi_{b2}(3P)$ mesons, from Table \ref{tab:Mass-shifts}] 
is in good accordance with the present experimental data: $M_{\chi_b(3P)} = 10.530 \pm 0.005 (\mbox{stat.}) \pm 0.009 (\mbox{syst.})$ GeV 
\cite{Aad:2011ih} and $M_{\chi_b(3P)} = 10.551 \pm 0.014 (\mbox{stat.}) \pm 0.017 (\mbox{syst.})$ GeV \cite{Abazov:2012gh}. 
It is interesting to observe that, in the case of the $\chi_b(3P)$ system, important threshold effects break the scheme for the splittings 
between $\chi_{b2} (3P) - \chi_{b1} (3P)$ and $\chi_{b1} (3P) - \chi_{b0} (3P)$ resonances, which holds in the $\chi_b(1P)$ and $\chi_b(2P)$ 
cases. 
See also Table \ref{tab:mass-bary-GI}, which reports results for the mass barycenters of $\chi_b(1P)$, $\chi_b(2P)$ and $\chi_b(3P)$ 
systems and the mass splittings between the members of the three multiplets, from our re-fit of the relativized QM of Sec. 
\ref{Strong decay widths}. Our results of Tables \ref{tab:mass-bary} and \ref{tab:mass-bary-GI} are substantially equivalent for the 
$\chi_b(1P)$ and $\chi_b(2P)$ systems, but differ for the $\chi_b(3P)$ multiplet, because of important threshold effects in the UQM case. 

We think that, when the high-statistics experiments at ATLAS and D0 are performed, they will be able to distinguish between quark model and 
unquenched quark model predictions for the masses of the states belonging to this multiplet. 
Indeed, our idea is that the QM can give a good reproduction of the experimental data, except in proximity to the thresholds, in which case 
the results should be corrected by unquenching the quark model.

Up to now, for the bottomonium case, we have investigated states that are far away from meson-meson decay thresholds, with the exception of 
$3P$ states, due to the complexity of the calculations as the values of the energies and the shells rise; nevertheless, the study of these 
higher excitations \cite{Brambilla:2010cs} will be the next step \cite{FerrettiandSantopinto}.

\subsection{Discussion of the results}
\label{Discussion of the results}
In this paper, we have studied higher bottomonia and provided results for the spectrum with self energy corrections and the open-bottom 
strong decays.
Specifically, the main results shown in this paper concern: 1) the first systematic calculation of the open-bottom strong decay widths of 
$b \bar b$ states within the $^3P_0$ model; 2) the systematic unquenching of a relativistic quark model for bottomonia, i.e. the development 
of an unquenched quark model for $b \bar b$ mesons; 3) the observation of the possible importance of continuum effects in the $\chi_b(3P)$ 
system. 

First of all, we computed the bottomonium spectrum with self energy corrections. 
In the UQM formalism of Refs. \cite{bottomonium,charmonium}, the effects of $q \bar q$ sea pairs are introduced explicitly into the QM 
through a QCD-inspired $^3P_0$ pair-creation mechanism.
The self energies we studied in this paper are corrections to the meson masses arising from the coupling to the meson-meson continuum. 
Neglected in naive QM's, these loop effects provide an indication of the quality of the quenched approximation used in QM calculations, 
where only valence quarks are taken into account. 
It is thus worthwhile seeing what happens when these pair-creation effects are introduced into the quark model, similarly to what is done 
in unquenched lattice QCD calculations \cite{Bali:1997am,Davies:1998im,Gray:2005ur,Burch:2007fj,Meinel:2010pv}.
Therefore, we could say that these kinds of studies can also be seen as inspections of the QM, of its power to predict the properties of 
hadrons and of its range of applicability: if the departure from QM results is important, one can see new physics emerging or better 
extra degrees of freedom.  

Several studies on the goodness of the quenched approximation in the QM have already been done, such as those of Refs. 
\cite{Bijker:2009up,Santopinto:2010zza,Bijker:2012zza,bottomonium,Geiger:1996re,Roberts:1992,Geiger-Isgur}. 
Many of them show that the quark model can predict several hadron properties with quite a high level of accuracy.
Nevertheless, there are observables whose expectation value on the valence component of a certain hadron is null, even if the expectation 
value on the sea component of the same hadron is nonzero: for example, this occurs in the case of the flavor asymmetry of the nucleon 
\cite{Santopinto:2010zza}, where one has to incorporate loop effects into the QM in order to carry out this kind of calculation. 
This is also the case of the self energy calculation of Ref. \cite{charmonium} and of the present paper.

Our results for the self energies of bottomonia show that the loop corrections to the spectrum of $b \bar b$ mesons (see Table 
\ref{tab:Mass-shifts}) are relatively small.  
Specifically for bottomonium states, they are approximately 1$-$2\% of the corresponding meson mass, while we have shown in Ref. 
\cite{charmonium} that the charmonium mass shifts induced by loop corrections are in the order of 2$-$6 \%. 
The relative mass shifts, namely the difference between the self energies of two meson states, are in the order of a few tens of MeV, but 
can become important in proximity to a threshold. 

In Ref. \cite{bottomonium}, a similar approach was already applied to lower bottomonia. However, the results of Ref. \cite{bottomonium} 
were only preliminary, not only because in the present paper the method is applied to the spectrum up to the excited states while in Ref. 
\cite{bottomonium} it was applied only to the low-lying ones. 
In Ref. \cite{bottomonium} the bare energies were taken as free parameters, and the sum of the bare and self energies was fitted 
to the experimental data. 
Thus, it was only a preliminary calculation and only now we are able to perform more complex coupled-channel calculations.
In the present paper, the bare energies were calculated within Godfrey and Isgur's relativized QM \cite{Godfrey:1985xj} and the sum of the 
bare and self energies was fitted to the experimental data. This is more consistent and much more elegan; above all, it increases the 
predictive power of the model: thus, we think that it constitutes an improvement to the previous calculation.
In addition, in Ref. \cite{bottomonium} the effective pair-creation strength $\gamma_0$ was fitted to the strong decay 
$\psi(3770) \rightarrow D \bar D$, calculated in SU$_{\mbox{f}}$(5), while in the present paper it is fitted to 
$\Upsilon(4S) \rightarrow B \bar B$. We think that this is more correct:
1) because a calculation of the properties of $b \bar b$ mesons based on a $^3P_0$-type model for the decay vertices has parameters that 
should be fitted in the most appropriate sector; thus the $b \bar b$ sector in this case, which considers the decay(s) of a $b \bar b$ 
meson(s); 2) because $b \bar b$ mesons have open-bottom decay thresholds that are located at high energies in comparison with the masses of 
the mesons belonging to the $c \bar c$ sector; i.e. the first $b \bar b$ meson decaying into a $B \bar B$ pair is a $4S$ one. 
Thus, if one wants to get reliable results for the open-bottom strong decays of higher bottomonia, one should fit the parameters in such a 
way is to get a good reproduction of the widths of these high radial (and orbital) excitations. In the charmonium case, the lowest energy 
state decaying into an open charm $D \bar D$ pair is the $\psi(3770)$, i.e. a $1D$ state. In general, $1D$ states lie at lower energies 
than $4S$ ones.

These continuum coupling effects are particularly important in the case of suspected non-$q \bar q$ states, such as the $X(3872)$ 
\cite{Choi:2003ue}. 
Indeed, it is true that, in general, the relativized QM \cite{Godfrey:1985xj} can provide a more precise overall description of the data. 
Nevertheless, as shown in Ref. \cite{charmonium} in the case of the $X(3872)$, the relativized QM may have problems when one considers 
states that are close to a meson-meson decay threshold. In this case, we think that it is necessary to introduce continuum coupling 
corrections.
At the moment, there are two possible interpretations for the $X(3872)$ \cite{Swanson:2006st}: a weakly bound $1^{++}$ $D \bar D^*$ 
molecule \cite{Hanhart:2007yq,Baru:2011rs,Aceti:2012cb} or a $c \bar c$ state 
\cite{Suzuki:2005ha,Meng:2007cx,charmonium,Danilkin:2010cc,Coito:2010if}, with $1^{++}$ quantum numbers. 
In particular, in Ref. \cite{charmonium} it is shown that the continuum coupling effects of the $X(3872)$ can give rise to $D\bar D^*$ and 
$D^*\bar D^*$ components in addition to the $c \bar c$ core and determine a downward energy shift, which is necessary to obtain a better 
reproduction of the experimental data.
This may also be the case of the $\chi_b(3P)$ resonances (or at least of one of them) that lie quite close to $B \bar B$, $B \bar B^*$ and 
$B^* \bar B^*$ decay thresholds. 
We think that the present experimental data \cite{Aad:2011ih,Abazov:2012gh} cannot exclude this possibility. 

Our work is not an extension of Godfrey and Isgur's relativized QM \cite{Godfrey:1985xj}, which we used to compute the bare energies of the 
$b \bar b$ states; this is the unquenching of the quark model \cite{charmonium}.
In Ref. \cite{Liu:2011yp}, the authors did something similar. However: 1) they used the non-relativistic potential model to compute 
the bare energies instead of Godfrey and Isgur's relativized model \cite{Godfrey:1985xj}, as we did; 2) they used the $^3P_0$ model to 
compute the self energy corrections, as we did, but they used a standard $^3P_0$ transition operator. Their results are therefore biased by 
the fact that they did not take the suppression of heavy quark pair-production into account, as we did. For example, in the case of 
$\chi_b(3P)$ states, their results are different from ours.

In a second stage, we also calculated the strong decay widths of $b \bar b$ states within a modified version of the $^3P_0$ model of hadron 
decays \cite{charmonium,bottomonium}. 
The corrections we introduced into the transition operator of the model include: 1) the use of a quark form factor, to take the effective 
size of the constituent quarks into account \cite{Geiger:1996re,Geiger-Isgur,SilvestreBrac:1991pw,charmonium,bottomonium}; 2) the 
replacement of the pair-creation strength $\gamma_0$ by the effective strength of Eq. (\ref{eqn:gamma0-eff}), which suppresses heavy 
$q \bar q$ pair-creation \cite{charmonium,Kalashnikova:2005ui,bottomonium}. 
The introduction of this effective mechanism is necessary, since in the original formulation of the $^3P_0$ model the flavor-independent 
pair-creation implies an unphysical enhancement of heavy-quark creation in comparison with light quark creation, without a fundamental 
reason for that \cite{Kalashnikova:2005ui}. 
These results for the strong decay widths, which required the re-fit of Godfrey and Isgur's mass formula to take the latest experimental 
data into account, may be particularly useful to experimentalists. Indeed, while knowledge of the $\Upsilon$ states and their decay 
modes is relatively good, this is not true for $\eta_b$ and $\chi_b$ states and, above all, for all the other states, such as the $D$, $F$ 
and $G$-wave ones, which we analysed in the present paper.

Our result for the width of the channel $\Upsilon(10860) \rightarrow B_s^* \bar B_s^*$, i.e. 1 MeV, is about 10 MeV too small in relation to the experimental value \cite{Nakamura:2010zzi}. 
The computed width is a function of the masses of the $\Upsilon(10860)$ and $B_s^*$ resonances, which, unfortunately, are still relatively poorly known.
New experiments will be important to determine the masses of the two resonances with higher precision.
Thus, we computed the width of the channel $\Upsilon(10860) \rightarrow B_s^* \bar B_s^*$ not only by taking into account the average values of the masses of the $\Upsilon(10860)$ and $B_s^*$ reported by the PDG, as it is done in Table \ref{tab:strong-decays}, but also by considering all the single experimental data for the masses of the two mesons quoted by the PDG \cite{Nakamura:2010zzi} and in the cited articles \cite{Aubert:2008ab,Lovelock:1985nb,Besson:1984bd,Chen:2008xia,Louvot:2008sc,Aquines:2006qg,Drutskoy:2006xc,Bonvicini:2005ci} (see Table \ref{tab:Y(10860)Bs*Bs*}). 
Our  theoretical results vary between 0 and 17 MeV, the average being 4.4 MeV. This is compatible with the experimental value, $\Gamma_{\Upsilon(10860) \rightarrow B_s \bar B_s^*} = 9.6_{-5.6}^{+7.1}$ MeV, within the experimental error.
\begin{table*}
\begin{tabular}{|c|c|c|c|c|}
\hline 
 & $M_{B_s^*} = 5416.4$ MeV \cite{Louvot:2008sc} & $M_{B_s^*} = 5411.7$ MeV \cite{Aquines:2006qg} & $M_{B_s^*} = 5418$ MeV \cite{Drutskoy:2006xc} & $M_{B_s^*} = 5414$ MeV \cite{Bonvicini:2005ci}\\    
\hline
$M_{\Upsilon(10860)} = 10876$ MeV \cite{Aubert:2008ab} &      2                                  &   0                                      &    3                                 &     0       \\
\hline
$M_{\Upsilon(10860)} = 10869$ MeV \cite{Aubert:2008ab} &      5                                  &   1                                      &    7                                 &     2       \\
\hline
$M_{\Upsilon(10860)} = 10845$ MeV \cite{Lovelock:1985nb} &      16                                &   15                                    &    13                               &  17        \\  
\hline  
$M_{\Upsilon(10860)} = 10868$ MeV \cite{Besson:1984bd} &      6                                  &   1                                      &    8                                 &     3       \\
\hline
$M_{\Upsilon(10860)} = 10879$ MeV \cite{Chen:2008xia} &      1                                  &   0                                      &    2                                 &     0       \\ 
\hline
$M_{\Upsilon(10860)} = 10888.4$ MeV \cite{Chen:2008xia} &      0                                  &   2                                      &    0                                 &     1       \\
\hline
\end{tabular}
\caption{Open-bottom strong decay width of the $\Upsilon(10860) \rightarrow B_s^* \bar B_s^*$ channel (in MeV), obtained through all the 
possible combinations of the masses of the $\Upsilon(10860)$ and $B_s^*$ mesons quoted by the PDG \cite{Nakamura:2010zzi}.
Our  theoretical results vary between 0 and 17 MeV, the average being 4.4 MeV, which is compatible with the experimental value, 
$\Gamma_{\Upsilon(10860) \rightarrow B_s \bar B_s^*} = 9.6_{-5.6}^{+7.1}$ MeV, within the experimental error. The width is computed within 
the $^3P_0$ model of Eqs. (\ref{eqn:3P0width}) and (\ref{eqn:relPSF}). We observe that some experiments report two different values for the 
mass of the $\Upsilon(10860)$ (see Refs. \cite{Aubert:2008ab} and \cite{Chen:2008xia}).}
\label{tab:Y(10860)Bs*Bs*}
\end{table*}
Our result for the total open-bottom width of the $\Upsilon(10860)$ has to be compared with only a fraction, of the order of $80\%$, of the total decay width of the meson ($55\pm28$ MeV \cite{Nakamura:2010zzi}) (see App. \ref{Y(10860)}). 
We observe that our theoretical result is compatible with the PDG average \cite{Nakamura:2010zzi}, being within the experimental error (see Table \ref{tab:open-bottom-exp}). 
This is a good result for a model with only one free parameter, i.e. $\gamma_0$. 

Our theoretical result for the $\Upsilon(11020)$ open-bottom decay width (see Table \ref{tab:strong-decays}) cannot be compared with 
the existing experimental data of the PDG \cite{Nakamura:2010zzi}. 
Indeed, the PDG reports the total width of the meson, $79\pm16$ MeV, but only gives the branching ratio 
$\Upsilon(11020) \rightarrow e^+ e^- = (1.6\pm0.5) \cdot 10^{-6}$ \cite{Nakamura:2010zzi}. Nevertheless, we can say that our theoretical 
result, i.e. 36 MeV, being smaller than the total decay width of the $\Upsilon(11020)$, at the moment is not incompatible with the present 
experimental data.

We think that the present paper may be a useful help to experimentalists in their search for new $b \bar b$ states. 
In the last few years, interest in heavy quarkonium physics has increased enormously, as has the number of collaborations devoted to 
the topic, because of the development of new B factories.
In particular, BaBar \cite{Prencipe:2012kb,Santoro:2012jq},  Belle \cite{Lange:2011qd}, CDF \cite{Yi:2009pz} and D0 \cite{Abazov:2012gh} 
have produced many interesting results. 
Moreover, all four detectors at LHC (Alice, Atlas, CMS and LHCb) have the capacity to study charmonia and bottomonia and have already 
produced some new results \cite{leonardo}, such as the discovery of a new $\chi_b(3P)$ system \cite{Aad:2011ih}. 
There are also approved proposals for new experiments, such as Belle II \cite{Yuan:2012jd}.
Therefore, we think it is important to study the properties of heavy mesons in order to provide (updated) information with spectra, strong 
decay widths, helicity amplitudes, and so on.

Finally, we would like to cite the interesting work by Hanhart {\it et al.} \cite{Hanhart:2007yq,Hanhart:2011jz}, who performed 
coupled-channel calculations with effective theories for the t-matrix in order to study the line shape of the $X(3872)$. 
In the future, we intend to do something similar, but at the quark level, in the case of the $\chi_b(3P)$ system. This will be the subject 
of a new paper \cite{FS-TBP}.
	
\begin{acknowledgments}
This work was supported by INFN, by Fondazione Angelo Della Riccia, Firenze, Italy, and by CONACYT (Grant No. 78833) and PAPIIT-DGAPA 
(Grant No. IN113711), Mexico.
\end{acknowledgments}

\begin{appendix}

\section{SU$_{\mbox{f}}$(5) couplings}
The SU$_{\mbox{f}}$(5) flavor couplings that we have to calculate in the $^3P_0$ model are 
$\langle F _{B}(14)F _{C}(32)|F _{A}(12)F _{0}(34)\rangle$ for the first diagram of Fig. \ref{fig:diagrammi3P0}, and 
$\langle F _{B}(32)F _{C}(14)|F _{A}(12)F _{0}(34)\rangle$ for the second diagram, in which $F_{X}(ij)$ represents the flavor wave function 
for the meson $X$ (i.e. the initial meson $A$, the final mesons $B$ and $C$ or the $^3P_0$ created pair $0$) made up of the quarks $i$ and $j$.
These overlaps can be easily calculated if we adopt a matrix representation of the mesons \cite{Ono:1983rd}. 
In this case, the two diagrams become, respectively,
\begin{equation}
	\begin{array}{rcl}
	\langle F _{B}(14)F _{C}(32)|F _{A}(12)F _{0}(34)\rangle & = & Tr[F _{A}F _{B}^{T}F _{0}F _{C}^{T}] \\ 
	& = & \frac{1}{\sqrt{5}}Tr[F _{A}F _{B}^{T}F _{C}^{T}] \mbox{ }, \\
	\langle F _{B}(32)F _{C}(14)|F _{A}(12)F _{0}(34)\rangle & = & Tr[F _{A}F _{C}^{T}F _{0}F _{B}^{T}] \\
	& = & \frac{1}{\sqrt{5}}Tr[F _{A}F _{C}^{T}F _{B}^{T}] \mbox{ }.
	\end{array}
\end{equation}

As an example, we calculate the flavor $\eta_b \rightarrow B^0 \bar B^0$ coupling.
The coupling in the $^3P_0$/UQM model can be written as
\begin{equation}
	\left\langle B(q_1 \bar q_4) C (q_3 \bar q_2) \right| \left. 
	A (q_1 \bar q_2) \Phi_0 (q_3 \bar q_4) \right\rangle \mbox{ },
\end{equation}
where $\Phi_0$ is the SU$_{\mbox{f}}$(5) flavor singlet
\begin{equation}
	\left| \Phi_0 \right\rangle = \frac{1}{\sqrt 5} \left( \left| u \bar u \right\rangle + \left| d \bar d \right\rangle 
	+ \left| s \bar s \right\rangle + \left| c \bar c \right\rangle 
	+ \left| b \bar b \right\rangle\right)  \mbox{ }.
\end{equation}
The states can be written as:
\begin{subequations}
\begin{equation}
	\left| \eta_b \right\rangle = \left| b \bar b \right\rangle  \mbox{ },
\end{equation}	
\begin{equation}
	\left| B^0 \right\rangle = \left| b \bar d \right\rangle  \mbox{ },
\end{equation}
\begin{equation}
	\left| \bar B^0 \right\rangle = \left| d \bar b \right\rangle  \mbox{ }.
\end{equation}
\end{subequations}
The flavor matrix element can then be written as the scalar product between $\left| \eta_b \Phi_0 \right\rangle$ and $\left| B^0 \bar B^0 \right\rangle$,
\begin{equation}
	\begin{array}{l}
		\left\langle B^0 \bar B^0 \right| \left. \eta_b \Phi_0 \right\rangle_{flavor} \\
		\hspace{0.5cm} = \left\langle b \bar d \right| \otimes \left\langle d \bar b \right| \\ 
		\hspace{1cm} \left| b \bar b \right\rangle \otimes \frac{1}{\sqrt 5} 
		\left( \left| u \bar u \right\rangle + \left| d \bar d \right\rangle 
	  + \left| s \bar s \right\rangle + \left| c \bar c \right\rangle 
		+ \left| b \bar b \right\rangle \right) \\ 
	  = \frac{1}{\sqrt 5} \left( \left\langle d \bar b b \bar d \right| \left. b \bar b d \bar d \right\rangle 
	   + \left\langle b \bar d d \bar b \right| \left. b \bar b d \bar d \right\rangle \right) 
		= \frac{1}{\sqrt 5} \mbox{ },
	\end{array}
\end{equation}
where the first contribution in parenthesis is zero and the second is one.

\section{Parameters of the $^3P_0$ pair-creation model}
\label{Model parameters}
The value of the width of the constituent quark form factor, $r_q=0.335$ fm, is taken from Ref. \cite{charmonium}.
The value of the harmonic oscillator parameter is taken as $\alpha=0.5$ GeV.
Finally, the value of the pair-creation strength, $\gamma_0^{\mbox{eff}}$ (see Table \ref{tab:parameters}), has to be fitted to the 
reproduction of experimental strong decay widths. 
We have chosen to fit $\gamma_0^{\mbox{eff}}$ to the experimental strong decay width $\Upsilon(4S) \rightarrow B \bar B$ 
\cite{Nakamura:2010zzi}.
In this case, since the created pair $q \bar q$ is $u \bar u$ or $d \bar d$, the effective pair-creation strength 
$\gamma_0^{\mbox{eff}}$ coincides with $\gamma_0$ [see Eq. (\ref{eqn:gamma0-eff})].
 
The decay width is calculated within the $^3P_0$ model \cite{SilvestreBrac:1991pw,Ackleh:1996yt} as 
\begin{equation}
	\begin{array}{rcl}
	\Gamma_{\Upsilon(4S) \rightarrow B \bar B} & = & 2 \Phi_{A \rightarrow BC} \left| \left\langle BC 
	\vec q_0  \, \ell J \right| T^\dag \left| A \right\rangle \right|^2  \\ 
	& = & 2 \Phi_{\Upsilon(4S) \rightarrow B \bar B} \\ 
	& & \left| \left\langle B \bar B 
	\vec q_0  \, 1 1 \right| T^\dag \left| \Upsilon(4S) \right\rangle \right|^2 \\ 
	& = & 21 \mbox{ MeV} \mbox{ },
	\end{array}
\end{equation}
where the factor of 2 is introduced since $\Upsilon(4S)$ decays into $B^0 \bar B^0$ or $B^+ B^-$, $\left\langle BC \vec q_0  \, \ell J \right| T^\dag \left| A \right\rangle$ is the $^3P_0$ amplitude describing the coupling between the meson $\left| A \right\rangle = \left| \Upsilon(4S) \right\rangle$ and the final state $\left| BC \right\rangle = \left| B \bar B \right\rangle$ and 
\begin{equation}
	\Phi_{A \rightarrow BC} = 2 \pi q_0 \frac{E_b E_c}{M_a}  
\end{equation}
is the standard relativistic phase space factor \cite{Ackleh:1996yt}, with $E_b = \sqrt{M_b^2 + q_0^2}$ and $E_c = \sqrt{M_c^2 + q_0^2}$. 

\section{Open-bottom and total decay widths of the $\Upsilon(10860)$}
\label{Y(10860)}
In this appendix, we show how we extracted the experimental result for the open-bottom decay width, $\Gamma_{\Upsilon(10860)}^{\mbox{exp.}} 
(\mbox{open bottom})$, and our theoretical result for the total width of the $\Upsilon(10860)$ resonance, 
$\Gamma_{\Upsilon(10860)}^{\mbox{th.}}$. 

The experimental result for the open-bottom decay width of the $\Upsilon(10860)$ \cite{Nakamura:2010zzi}
\begin{equation}
	\Gamma_{\Upsilon(10860)}^{\mbox{exp.}} (\mbox{open bottom}) = 42^{+29}_{-24} \mbox{MeV}
\end{equation}  
is extracted in the following way: 
1) the PDG average for the total width of the $\Upsilon(10860)$, i.e. 55 MeV \cite{Nakamura:2010zzi}, is multiplied by a factor of 0.7664. 
This is the branching ratio corresponding to the open-bottom strong decays and is the sum of the following contributions: $5.5\%$ 
($B \bar B$), $13.7\%$ ($B \bar B^*$ and $\bar B  B^*$), $38.1\%$ ($B^* \bar B^*$), $0.5\%$ ($B_s \bar B_s$), $1.34\%$ ($B_s \bar Bs^*$ and 
$\bar B_s  Bs^*$) and $17.5\%$ ($B_s^* \bar B_s^*$) \cite{Nakamura:2010zzi}; 2) the lower and upper extremes of the interval 
$42^{+29}_{-24}$ MeV, 18 and 71 MeV, are obtained by multiplying the lower and upper extremes of the interval $55\pm28$ MeV [the total 
width of the $\Upsilon(10860)$], 27 and 83 MeV, by the branching fractions $67.2\%$ and $86.1\%$, respectively \cite{Nakamura:2010zzi}; 
3) these branching fractions, $67.2\%$ and $86.1\%$, are obtained by summing the lower or upper limits, respectively, of the branching 
ratios of each open-bottom channel: $B \bar B$, $B \bar B^*$, and so on. For example, the lower limit of the branching fraction for the 
channel $\Upsilon(10860) \rightarrow B \bar B$ is $4.5\%$ and its upper limit is $6.5\%$ \cite{Nakamura:2010zzi}.

$\Gamma_{\Upsilon(10860)}^{\mbox{th.}}$ is extrapolated in the following way:
1) the theoretical open-bottom width of the $\Upsilon(10860)$, i.e. 71 MeV (see Table \ref{tab:open-bottom-exp}), is divided by the 
experimental open-bottom branching fraction, i.e. $76.64\%$. This yields 93 MeV; 2) to estimate the error on the theoretical result, one has 
to divide the theoretical open-bottom width, 71 MeV, by the lower and upper extremes of the experimental interval for the open-bottom 
branching fraction, $67.2\%$ and $86.1\%$, respectively. The procedure for extracting these numbers, $67.2\%$ and $86.1\%$, has already been 
explained in the first part of the appendix. In this way, one gets the extremes of the theoretical interval for the total width of the 
$\Upsilon(10860)$: 82 and 106 MeV; 3) now, one can write the theoretical result for the total width of the $\Upsilon(10860)$, with its 
error: 
\begin{equation}
	\Gamma_{\Upsilon(10860)}^{\mbox{th.}}  = 93_{-11}^{+13} \mbox{MeV} \mbox{ }.
\end{equation}
As in the case of the open-bottom width of the $\Upsilon(10860)$, our theoretical result $\Gamma_{\Upsilon(10860)}^{\mbox{th.}}$ is 
compatible with the experimental data \cite{Nakamura:2010zzi} within the experimental error.

\end{appendix}


\end{document}